\documentclass{article}
\usepackage{geometry}
\usepackage{graphicx}

\usepackage{hyperref}
\hypersetup{colorlinks=true,linkcolor=blue,citecolor=blue}
\usepackage{cite}

\begin{document}

\title{\textbf{Optimal probabilistic dense coding schemes}}


\author{Roger A. K\"ogler\footnote{E-mail: rogerkogler@gmail.com}\, and Leonardo Neves \vspace{1mm} \\
\textit{\normalsize{ Departamento de F\'isica, Universidade Federal de Minas Gerais}} \\
 \textit{\normalsize{Belo Horizonte, MG 31270-901, Brazil}}}

\date{}

\maketitle

\begin{abstract}
Dense coding with non-maximally entangled states has been investigated in many different scenarios. We revisit this problem for protocols adopting the standard encoding scheme. In this case, the set of possible classical messages cannot be perfectly distinguished due to the non-orthogonality of the quantum states carrying them. So far, the decoding process has been approached in two ways:  (i) The message is always inferred, but with an associated (minimum) error; (ii) the message is inferred without error, but only sometimes; in case of failure, nothing else is done. Here, we generalize on these approaches and propose novel optimal probabilistic decoding schemes. The first uses quantum-state separation to increase the distinguishability of the messages with an optimal success probability. This scheme is shown to include (i) and (ii) as special cases and continuously interpolate between them, which enables the decoder to trade-off between the level of confidence desired to identify the received messages and the success probability for doing so. The second scheme, called multistage decoding, applies \emph{only} for qudits ($d$-level quantum systems with $d>2$) and consists of further attempts in the state identification process in case of failure in the first one.  We show that this scheme is advantageous over (ii) as it increases the mutual information between the sender and receiver. \\

\end{abstract}

\section{Introduction}
\label{intro}
Quantum information theory has impacted in many ways our knowledge on how information may be stored, manipulated, and transmitted \cite{BarnettBook}. Much of its achievements were consequence of a unique type of correlation among physical systems exhibited only by the quantum systems, namely the entanglement. This is, for instance, the resource that enables information to be transmitted in an unparalleled and secure fashion as shown in quantum teleportation \cite{Bennett93} and quantum cryptography \cite{Ekert91}, respectively. Another remarkable application of entanglement is the quantum communication protocol called dense coding, proposed by Bennett and Wiesner \cite{Bennett92}. They have shown that if two parts share this resource, one part can send more information to another than would be possible without entanglement. The simplest example of this protocol works as follows: Alice, the sender, and Bob, the receiver, share a pair of qubits (two-level systems) in a known maximally entangled Bell state. She performs, on her qubit, one of the four unitaries $\hat{I}$, $\hat{\sigma}_X$, $\hat{\sigma}_Y$, and $\hat{\sigma}_Z$ given by the identity and the Pauli operators, respectively. Each of these local unitary operations maps the initial two-qubit state into one of the four orthogonal Bell states, which then can be used to encode two bits of information. After this encoding, Alice sends her qubit to Bob who can extract these two bits of information by performing a Bell-state measurement on both qubits in his possession. Therefore, by sending a single qubit, part of a shared maximally entangled pair, Alice is able to communicate two bits of classical information to Bob. If there was no entanglement, a single qubit would enable just one bit of information to be transmitted between them.
 
Likewise any other quantum information protocol that relies on maximally entangled states to work perfectly, dense coding is also affected when the entanglement shared between Alice and Bob is not maximal. In this scenario, many schemes have been proposed to optimize the protocol 
\cite{Barenco95,Hausladen96,Bose00,Hao00,Ziman00,Hiroshima01,Bowen01,Pati05,Mozes05,Ji06,Wu06,Bourdon08,Beran08}. We can divide these schemes into two main lines of investigation. In the first, one optimizes the encoding/decoding processes while trying to transmit the same amount of information that would be possible with a maximally entangled state \cite{Barenco95,Hausladen96,Bose00,Hao00,Ziman00,Hiroshima01,Bowen01,Pati05}. In the second, one seeks to maximize the number of perfectly distinguishable messages that can be encoded and transmitted through the shared entangled state \cite{Mozes05,Ji06,Wu06,Bourdon08,Beran08}.\footnote{For qudits ($d$-level systems), this number will be intermediate between the one achieved by a maximally entangled and a non-entangled state. For qubits, this number would be 3, but as shown in \cite{Mozes05}, it is impossible to encode three perfectly distinguishable messages in any partially entangled two-qubit state.} Our work follows the first line of investigation and, in this context, we assume that Alice's encoding scheme will be the standard one presented in the original dense coding proposal \cite{Bennett92}. This encoding, performed with a set of orthogonal unitary operators\footnote{To be defined in Sect.~\ref{sec:QDC}.} applied with equal \emph{a priori} probabilities, maximizes the information transmission capacity of the protocol, as shown in Refs.~\cite{Bose00,Hiroshima01,Brub04}. 

Within the framework established above, the set of possible classical messages encoded by Alice will not be perfectly distinguishable due to the non-orthogonality of the quantum states carrying them. What remains is to optimize Bob's decoding process, i.e., to find the measurement strategy which optimally discriminates among these non-orthogonal states. So far, this problem has been approached in two ways: 
\begin{itemize} 
\item[(i)] Bob implements a measurement which \emph{always} infers Alice's message and minimizes the inevitable error in doing so. This strategy admits no failure probability and, as shown in \cite{Barenco95,Hausladen96}, it maximizes the mutual information between them in the dense coding.
 
\item[(ii)] Bob implements a measurement that infers Alice's message \emph{without} error, but only with a certain success probability, which must be maximized in order to optimize the process. In case of failure, nothing else is done. As shown in \cite{Hao00,Pati05}, this strategy provides full confidence in identifying the non-perfectly distinguishable messages at the cost of reducing the mutual information between them in comparison with the case (i). 

\end{itemize}
The decoding processes (i) and (ii) are implemented via minimum-error (ME) and maximum-confidence (MC) measurements, respectively. Reviews on these  discrimination strategies may be found in Refs.~\cite{Chefles00,Barnett09,Bergou10}. The field of quantum-state discrimination has rapidly evolved in the last years, becoming a fundamental tool in quantum information theory. Many new optimized measurement strategies have been proposed \cite{Hayashi08,Jimenez11,Bagan12,Zhang14,Prosser16} and applied in protocols like, for instance, quantum teleportation \cite{Neves12} and entanglement swapping \cite{Prosser14}. 

Here, we propose the application of the probabilistic discrimination strategies introduced in Refs.~\cite{Jimenez11,Prosser16} by one of us and co-workers, in the decoding process of the dense coding protocol. Firstly, we present a family of probabilistic decoding schemes which include (i) and (ii) described above as special cases and continuously interpolate between them. It is implemented via quantum-state separation, namely a map that transforms the states carrying part of Alice's message into more distinguishable states, with an optimal success probability \cite{Prosser16}. After a successful event, the decoding is accomplished with a ME measurement. This scheme allows Bob to trade-off between the level of confidence he wishes to identify Alice's messages and the success probability for this task. The second group of probabilistic decoding schemes applies \emph{only} for dense coding with qudits. It relies on the following fact: After a failed discrimination attempt in a probabilistic strategy, the qudit states, in general, still carry information about the input states.  Thus, they can be subjected to further discrimination attempts and the process may be iterated \cite{Jimenez11}. We call this process multistage decoding and show that it improves over scheme (ii) increasing the mutual information between Alice and Bob. 

This paper is organized as follows. In Sect.~\ref{sec:QDC} we describe the encoding/decoding processes in the dense coding studied here and provide a simple expression for the mutual information, the figure of merit to be optimized in our protocols. In Sect.~\ref{sec:ME} we briefly review the decoding scheme (i), which is used as the final step in our protocols. In Sect.~\ref{sec:separation} we introduce the family of probabilistic protocols assisted by quantum-state separation. The multistage decoding is presented and discussed in Sect.~\ref{sec:multi-stage}. In Sect.~\ref{sec:QKD} we present a qualitative analysis of the application of these decoding schemes in quantum key distribution. Finally, we conclude the paper in Sect.~\ref{sec:conc}.

\section{Dense coding}
\label{sec:QDC}

\subsection{Encoding}
\label{sec:enc}

Let us consider Alice and Bob sharing a \emph{known} bipartite entangled state $|\Psi\rangle_{12}$ where system 1 (2) lives in a $d_{1(2)}$-dimensional Hilbert space, $\mathcal{H}_1$ ($\mathcal{H}_2$), and it is in Bob's (Alice's) hands. Using the Schmidt decomposition \cite{BarnettBook}, this state can be written as
\begin{equation}    \label{eq:psi00}
|\Psi\rangle_{12}=\sum_{l=0}^{D-1}a_l|l\rangle_1|l\rangle_2,
\end{equation}
where $a_l$'s are strictly positive real numbers satisfying $\sum_la_l^2=1$ and, hence, $D$ is the Schmidt rank of the state, which satisfies $D\leq\min(d_1,d_2)$ (see Table~\ref{tab:d} for a prompt reference on these quantities). The Schmidt bases $\{|l\rangle_\ell\}_{l=0}^{d_\ell-1}$ are assumed to coincide with the computational basis spanning $\mathcal{H}_\ell$ ($\ell=1,2$). If that is not the case in the beginning, they can apply local unitary rotations to align them.

\begin{table}
\begin{center}
\caption{Dimension of the Hilbert spaces and Schmidt rank of the entangled state (\ref{eq:psi00}).}
\label{tab:d}      
\begin{tabular}{lll}
\hline\noalign{\smallskip}
$d_1$ & $d_2$ &  $D$ \\
\noalign{\smallskip}\hline\noalign{\smallskip}
$\dim(\mathcal{H}_1)\;\;\;$ & $\dim(\mathcal{H}_2)\;\;\;$ &  Schmidt rank $(\leq \min(d_1,d_2))$ \\
\noalign{\smallskip}\hline
\end{tabular}
\end{center}
\end{table}

\begin{figure}[tbp]
\centerline{\includegraphics[width=1\textwidth]{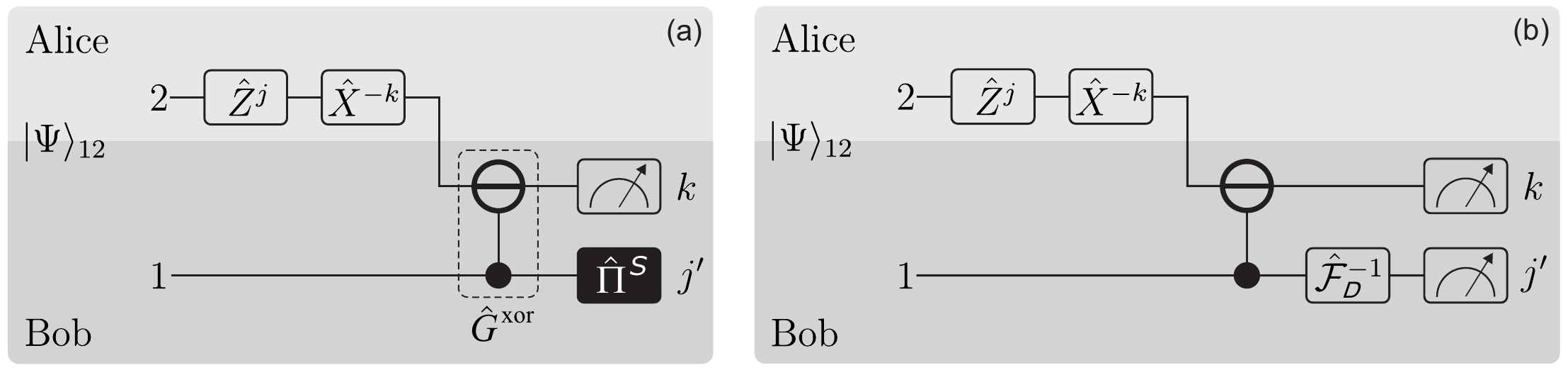}}
\caption{\label{fig:circs1} \textbf{a} General form of the circuit for the dense coding protocols studied here. The operation $\hat{\Pi}^\mathcal{S}$ in the black box will take different forms according to Bob's measurement strategy. \textbf{b} Circuit for the standard dense coding \cite{Bennett92} which also represents the optimal protocol for non-maximally entangled states when no failure probability is admitted \cite{Barenco95,Hausladen96} (see Sect.~\ref{sec:ME}). The measurement outcome $j'$ emphasizes that errors may occur when inferring the part of the message encoded in system 1.}
\end{figure}

In the dense coding protocol studied here, Alice encodes one of $\mathcal{N}=d_2D$ messages with the same \emph{a priori} probability ($1/\mathcal{N}$), by performing a local unitary operation $\hat{X}_2^{-k}\hat{Z}_2^{j}$ (for $k=0,\ldots,d_2-1$ and $j=0,\ldots,D-1$) on her system. Thus, the classical message encoded in the quantum state will be defined by the ordered pair $(j,k)$. This is illustrated in the circuits of Fig.~\ref{fig:circs1}. The unitaries $\hat{X}_2$ and $\hat{Z}_2$ are the generalized Pauli operators acting on a $d_2$- and $D$-dimensional Hilbert space, respectively. They are defined by their action on the computational basis as
\begin{equation}   \label{eq:opX}
\hat{X}|l\rangle=|l\oplus 1\rangle,
\end{equation}
and
\begin{equation}   \label{eq:opZ}
\hat{Z}|l\rangle=e^{2\pi i l/D}|l\rangle,
\end{equation}
where $\oplus$ denotes addition modulo $d_2$. With this encoding process, the state shared between Alice and Bob in Eq.~(\ref{eq:psi00}) is transformed into
\begin{eqnarray}
|\Psi_{jk}\rangle_{12} & = & \hat{X}_2^{-k}\hat{Z}_2^{j}|\Psi\rangle_{12}\nonumber\\ 
& = & \sum_{l=0}^{D-1}a_le^{2\pi i jl/D}|l\rangle_1|l\ominus k\rangle_2\nonumber\\
& = & \hat{G}_{12}^{\rm xor}|\alpha_j\rangle_1|k\rangle_2.   \label{eq:psijk}
\end{eqnarray}
The unitary and hermitian operator $\hat{G}_{12}^{\rm xor}$ is a generalization of the controlled-\textsc{not} gate for arbitrary dimensions \cite{Alber01}. Having system 1 (2) as the control (target), its action on the computational basis is defined by\footnote{Even if the bases $\{|m\rangle_1\}$ and $\{|n\rangle_2\}$  have different cardinalities, the $\hat{G}_{12}^{\rm xor}$ gate can still be defined, as pointed out in \cite{Daboul03}.} 
\begin{equation}
\hat{G}_{12}^{\rm xor}|m\rangle_1|n\rangle_2=|m\rangle_1|m\ominus n\rangle_2,
\end{equation}
where $\ominus$ denotes subtraction modulo $d_2$. The state of system 1 in Eq.~(\ref{eq:psijk}) is given by
\begin{equation}    \label{eq:sym}
|\alpha_j\rangle_1=\sum_{l=0}^{D-1}a_le^{2\pi i jl/D}|l\rangle_1=\hat{Z}_1^j\sum_{l=0}^{D-1}a_l|l\rangle_1,
\end{equation}
and the set $\{|\alpha_j\rangle_1\}_{j=0}^{D-1}$ is composed by {\it equally likely symmetric states}. They are equally likely because the encoding is performed with the same \emph{a priori} probability. In addition, they are symmetric under the action of $\hat{Z}$ because the conditions $|\alpha_j\rangle=\hat{Z}|\alpha_{j-1}\rangle=\hat{Z}^j|\alpha_0\rangle$ and $\hat{Z}|\alpha_{D-1}\rangle=|\alpha_0\rangle$ apply \cite{Chefles98}.

As pointed out in Ref.~\cite{Mozes05}, in the encoding given by Eq.~(\ref{eq:psijk}), the unitary $\hat{X}_2$ [Eq.~(\ref{eq:opX})] is the ``classical'' part of the process since it enables one to encode one of $d_2$ perfectly distinguishable (PD) messages in a $d_2$-level qudit. This is no better than using a ``$d_2$-state'' classical communication channel. On the other hand, $\hat{Z}_2$ [Eq.~(\ref{eq:opZ})] is the quantum enhancement, as it introduces, locally, relative phases on the shared entangled state allowing one to encode (in combination with $\hat{X}_2$) one of $d_2D$ messages, which may or may not be PD, thus increasing the information capacity over any $d_2$-state classical channel. The extent of this enhancement, however, depends on the initial entangled state shared between Alice and Bob. 
If it is \emph{uniform}, i.e., if  $a_j=1/\sqrt{D}$ $\,\forall\, j$ in Eq.~(\ref{eq:psi00}), the states $|\Psi_{jk}\rangle_{12}$ in Eq.~(\ref{eq:psijk}) become
\begin{equation}    \label{eq:PDstate}
|\Psi_{jk}\rangle_{12}=\hat{G}_{12}^{\rm xor}\hat{\mathcal{F}}_{1,D}|j\rangle_1|k\rangle_2,
\end{equation}
where $\hat{\mathcal{F}}_{1,D}$ is the discrete Fourier transform acting on the $D$-dimensional (sub)space\footnote{The Fourier transform is defined as $\hat{\mathcal{F}}_{1,D}=D^{-1/2}\sum_{m,n=0}^{D-1}e^{2\pi imn/D}|m\rangle\langle n|$. Along the paper, the (sub)space where it acts depends on the relationship between $d_1$, $d_2$, and $D$ (see Table~\ref{tab:d}). If $d_1>d_2$, then $D\leq d_2$ so that $\hat{\mathcal{F}}_{1,D}$ acts, necessarily, on a $d_2$-dimensional subspace of $\mathcal{H}_1$. If $d_1<d_2$, then $D\leq d_1$ so that $\hat{\mathcal{F}}_{1,D}$ acts on a subspace of $\mathcal{H}_1$, for $D<d_1$, or the entire $\mathcal{H}_1$ space, for $D=d_1$.} of system 1. In this case is easy to check that
\begin{equation}
\langle\Psi_{mn}|\Psi_{jk}\rangle=\delta_{jm}\delta_{kn},
\end{equation}
$\forall\, j,m=0,\ldots,D-1$ and $\forall\, k,n=0,\ldots,d_2-1$, which means that Alice can encode $d_2D$ PD messages. In addition, if $D=\min(d_1,d_2)$, the entanglement of (\ref{eq:psi00}) is maximal and the number of possible PD messages is the maximum allowed for the quantum channel, that is, $\mathcal{N}_{\rm PD}=d_2^2$ if $d_1>d_2$, and $\mathcal{N}_{\rm PD}=d_1d_2$ if $d_1<d_2$. However, if the shared entangled state is not uniform (which implies partial entanglement), the overlap between the states $|\Psi_{jk}\rangle_{12}$ in Eq.~(\ref{eq:psijk}) will be 
\begin{equation}
\langle\Psi_{mn}|\Psi_{jk}\rangle=\underbrace{\langle\alpha_m|\alpha_j\rangle}_{\neq\delta_{jm}}\delta_{kn},
\end{equation}
meaning that the possible messages Alice can encode will not be PD, as a consequence of the non-orthogonality of the symmetric states $|\alpha_j\rangle$ defined in (\ref{eq:sym}).

\subsection{Decoding}
After receiving system 1 from Alice, Bob performs a joint measurement on 1 and 2 to determine the corresponding state $|\Psi_{jk}\rangle_{12}$ and, hence, decode her message.  First, he applies the unitary and hermitian   $\hat{G}_{12}^{\rm xor}$ gate. From Eq.~(\ref{eq:psijk}), this yields  
\begin{equation}   \label{eq:Message_state}
\hat{G}_{12}^{\rm xor}|\Psi_{jk}\rangle_{12}=|\alpha_j\rangle_1|k\rangle_2.
\end{equation}
Now, the state of system 2 can be perfectly determined by a projective measurement onto the computational basis. To finish the decoding, Bob must identify the state of system 1 [Eq.~(\ref{eq:sym})] through a proper measurement, which is the problem we will address here. The process described so far is sketched in the circuit of Fig.~\ref{fig:circs1}a, where the operation $\hat{\Pi}^\mathcal{S}$ in the black box will take different forms according to Bob's measurement strategy.

As shown in Eq.~(\ref{eq:PDstate}), if the state shared between Alice and Bob is uniform, the symmetric state in (\ref{eq:sym}) reduces to $|\alpha_j\rangle_1=\hat{\mathcal{F}}_{1,D}|j\rangle_1$. Therefore, the set $\{|\alpha_j\rangle_1\}_{j=0}^{D-1}$ is composed by mutually orthogonal states which can be perfectly discriminated by applying an inverse Fourier transform followed by a projective measurement onto the computational basis, as sketched in Fig.~\ref{fig:circs1}b. This is the standard dense coding protocol devised by Bennett and Wiesner in Ref.~\cite{Bennett92}. 

For a non-maximally and non-uniform entangled state, the set $\{|\alpha_j\rangle_1\}_{j=0}^{D-1}$ is composed by non-orthogonal states, for which there is no measurement able to identify among them deterministically \emph{and} with full confidence. Therefore, the optimal decoding process requires optimal discrimination among those states. In the next sections we shall present measurement strategies applied to this problem, focusing our attention on the probabilistic ones.

\subsection{Mutual information in the dense coding}

Before addressing the optimal decoding strategies, we briefly describe the mutual information between Alice and Bob in the dense coding discussed above. This will be our figure of merit to analyze the performance of the protocols. 

The mutual information between two random variables $M$ and $R$ quantifies the amount of information about $M$ acquired by determining the value of $R$ \cite{CoverBook}. In the present context, $M$ take values in the set of possible messages encoded by Alice and $R$ in the measurement results of Bob's decoding. As shown in Appendix, this quantity can be written in a simplified and more instructive form as 
\begin{equation}   \label{eq:IM_general}
[I(M;R)]^{\mathcal{S}}=\log_2d_2D-[H(M_1|R_1)]^{\mathcal{S}},
\end{equation}
where 
\begin{equation}   \label{eq:Entrop_cond}
[H(M_1|R_1)]^{\mathcal{S}}=-\frac{1}{D}\sum_{j=0}^{D-1}\sum_{l=0}^{N-1}{\rm Tr}(\hat{\rho}_j\hat{\Pi}_l^{\mathcal{S}})\log_2{\rm Tr}(\hat{\rho}_j\hat{\Pi}_l^{\mathcal{S}}).
\end{equation}
In these expressions, $\mathcal{S}$ denotes the discrimination strategy to be adopted by Bob to identify the state of system 1. $[H(M_1|R_1)]^{\mathcal{S}}$ is the conditional entropy which quantifies the uncertainty about the random variable $M_1$ when we know the value of $R_1$. Both variables now are associated with the encoding/decoding process for the part of the message encoded in system 1, i.e., $M_1$ take values in the set $\{\hat{\rho}_j=|\alpha_j\rangle\langle\alpha_j|\}_{j=0}^{D-1}$ of symmetric states (\ref{eq:sym}), and $R_1$ in the set of outcomes $\{\omega_l\}_{l=0}^{N-1}$ corresponding to the generalized measurement $\{\hat{\Pi}_l^{\mathcal{S}}\}_{l=0}^{N-1}$. In (\ref{eq:Entrop_cond}), ${\rm Tr}(\hat{\rho}_j\hat{\Pi}_l^{\mathcal{S}})$ gives the probability of obtaining the outcome $\omega_l$ given that the prepared state was $\hat{\rho}_j$. Note that the number of measurement operators, $N$, is arbitrary, as long as they form a physically realizable measurement. Here, we shall consider only $N\geq D$.

The conditional entropy in Eq.~(\ref{eq:Entrop_cond}) is associated with the uncertainties in the discrimination of the symmetric states (\ref{eq:sym}). It obeys $0\leq [H(M_1|R_1)]^{\mathcal{S}}\leq\log_2D$, where 0 and $\log_2D$ indicates no uncertainty (orthogonal states) and maximum uncertainty (identical states), respectively. Thus, from Eq.~(\ref{eq:IM_general}), the mutual information in the dense coding satisfies 
\begin{equation}
\log_2d_2\leq [I(M;R)]^{\mathcal{S}}\leq\log_2d_2D,
\end{equation} 
where the minimum corresponds to a state with no entanglement and the maximum to a uniform entangled state. For non-uniform entangled states, we must find optimal discrimination strategies which minimize $[H(M_1|R_1)]^{\mathcal{S}}$, as will be shown next.

\section{Decoding with minimum-error quantum measurements}
\label{sec:ME}

Dense coding with non-maximally entangled states, using the encoding scheme discussed in Sect.~\ref{sec:enc}, has been first studied by Barenco and Ekert  \cite{Barenco95}. Restricting to the case of qubits, they have shown that the standard decoding process [see Fig.~\ref{fig:circs1}b] presented in Ref.~\cite{Bennett92} was already the optimal one, maximizing the mutual information between Alice and Bob. Later, this result was generalized for arbitrary dimensions \cite{Hausladen96}. In the standard process, Bob determines the state of system 1 with the projective measurement
\begin{eqnarray}   \label{eq:Pi_ME}
\hat{\Pi}^{\rm ME}_j&=&\hat{\mathcal{F}}_{1,D}|j\rangle\langle j|\hat{\mathcal{F}}_{1,D}^{-1} \nonumber\\
&=&|\mu_j\rangle\langle\mu_j|,
\end{eqnarray}
for $j=0,\ldots,D-1$. This is exactly the optimal measurement that discriminates among equally likely symmetric states with minimum error \cite{Ban97}. Therefore, by applying the ME strategy in the decoding, Bob performs the standard protocol which is the optimal one when no failure probability is admitted in the process. Using Eqs.~(\ref{eq:sym}), (\ref{eq:IM_general}), (\ref{eq:Entrop_cond}), and (\ref{eq:Pi_ME}), the mutual information between Alice and Bob will be
 \begin{equation}   \label{eq:IM_ME}
[I(M;R)]^{\rm ME} = \log_2 d_2D + \frac{1}{D} \sum_{j,l=0}^{D-1} p_{jl} \log_2  p_{jl} ,
\end{equation}
where
\begin{eqnarray}   
p_{jl}&=&|\langle \mu_l|\alpha_j\rangle|^2 \nonumber\\
&=&\frac{1}{D}\sum_{m,n=0}^{D-1}a_ma_n\exp[2\pi i(m-n)(j-l)/D]
\label{eq:Pjl_EM}
\end{eqnarray}
is the probability of obtaining an outcome $\omega_l$ if the state of system 1 is $|\alpha_j\rangle$. It is easy to see that if the entangled state is uniform, $p_{jl}=\delta_{jl}$ and $[I(M;R)]^{\rm ME} = \log_2 d_2D$. Also, if there is no entanglement, $p_{jl}=1/D$ and $[I(M;R)]^{\rm ME} = \log_2 d_2$, as expected. For any other entangled state $[I(M;R)]^{\rm ME}>\log_2 d_2$, showing that even partial entanglement is always better than no entanglement at all.

\begin{figure}[tbp]
\centerline{\includegraphics[width=.7\textwidth]{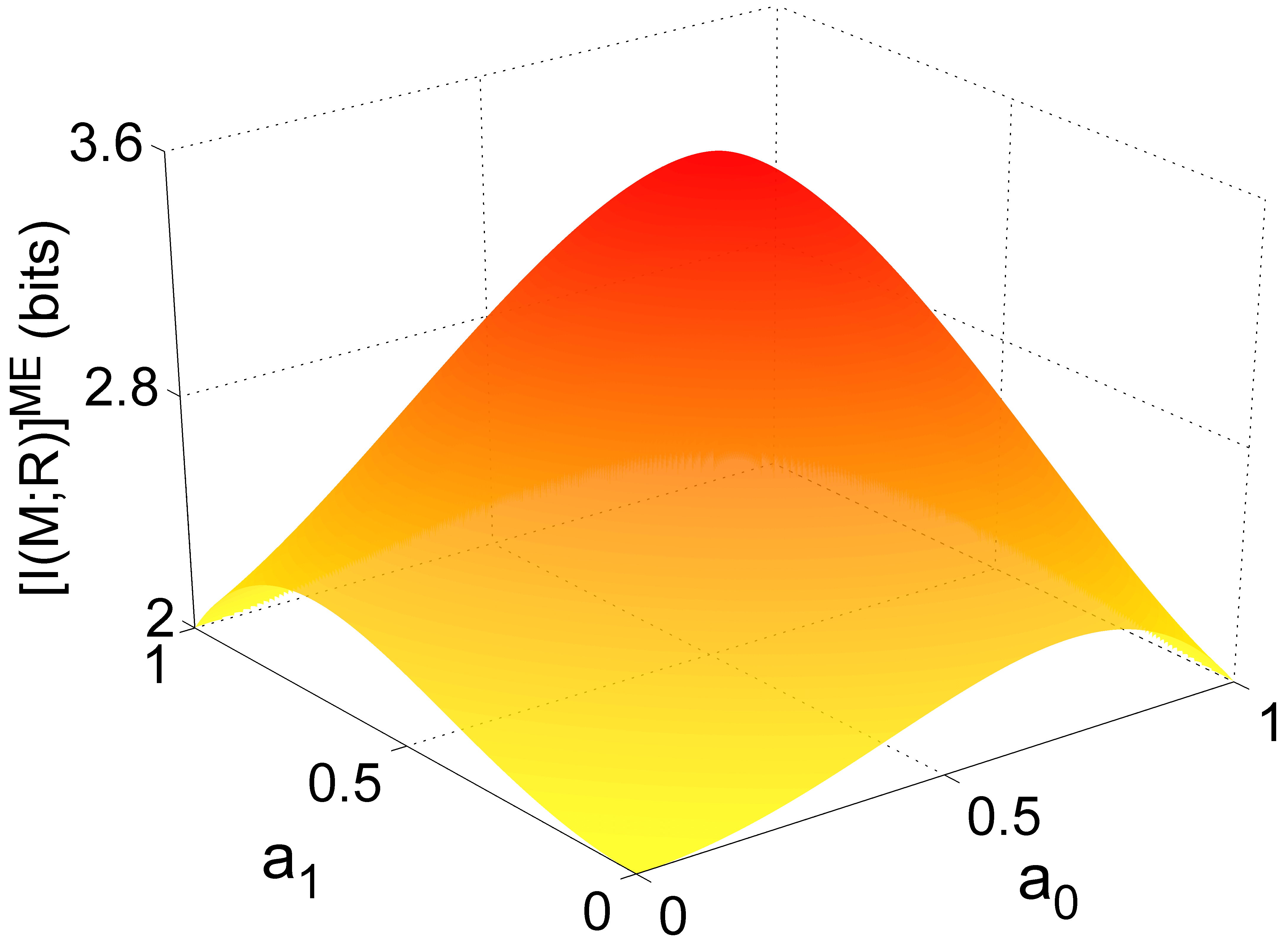}}
\caption{\label{fig:IM_ME} Mutual information in the dense coding as a function of the Schmidt coefficients $a_0$ and $a_1$, for an entangled state with $d_1=3$ and $d_2=4$. This plot corresponds to the decoding performed with ME measurements and is obtained from Eqs.~(\ref{eq:IM_ME}) and (\ref{eq:Pjl_EM}).}
\end{figure}

As an example, consider an entangled state with $d_1=3$ and $d_2=4$. Using Eqs.~(\ref{eq:IM_ME}) and (\ref{eq:Pjl_EM}) we plot the corresponding maximal mutual information as a function of the Schmidt coefficients $a_0$ and $a_1$ ($a_2=\sqrt{1-a_0^2-a_1^2}$) in Fig.~\ref{fig:IM_ME}. The maximum, $[I(M;R)]^{\rm ME} = \log_212$, is reached for a maximally entangled state; the minimum values, 2 bits, correspond to product states ($a_0=a_1=0$, $a_0=a_2=0$, and $a_1=a_2=0$). The curves in the axes $a_0=0$, $a_1=0$, and $a_2=0$, correspond to entangled states with non-maximal Schmidt rank ($D=2$), but with Alice encoding the same number of messages of a full rank state. Obviously, these messages will be even less distinguishable as the states carrying them are also linearly dependent, thus reducing the mutual information. This situation differs from the one presented in Sect.~\ref{sec:QDC} and will not be addressed here.

\section{Decoding assisted by quantum-state separation}
\label{sec:separation}

The first probabilistic approach to the dense coding with non-maximally entangled states (considering the encoding presented in Sect.~\ref{sec:enc}) was given by Hao \emph{et al.} for the case of qubits \cite{Hao00} and later generalized by Pati \emph{et al.} \cite{Pati05} for the case of qudits. These authors have shown that with this non-ideal resource Alice could communicate $\log_2d_2D$ classical bits to Bob---the maximum achieved, deterministically, by a maximally entangled state---with a certain success probability. 

Here, we introduce a new family of probabilistic dense coding schemes which have those in Refs.~\cite{Hao00,Pati05} as special cases and interpolate continuously between them and the ones presented in \cite{Barenco95,Hausladen96} (described in the previous section). Our approach is based on Bob's ability of mapping the symmetric states $|\alpha_j\rangle$ given by Eq.~(\ref{eq:sym}) into the symmetric states $|\beta_j(\xi)\rangle$ given by
\begin{equation}   \label{eq:beta_st}
|\beta_j(\xi)\rangle=\sum_{l=0}^{D-1}b_l(\xi)e^{2\pi ijl/D}|l\rangle,
\end{equation}
where
\begin{equation}   \label{eq:b_xi}
b_l(\xi)=\left[(1-\xi)a_l^2+\frac{\xi}{D}\right]^{1/2}
\end{equation}
and $\xi\in[0,1]$. It is easy to see that $|\beta_j(0)\rangle=|\alpha_j\rangle$ and 
\begin{eqnarray}
|\beta_j(1)\rangle&=&\frac{1}{\sqrt{D}}\sum_{l=0}^{D-1}e^{2\pi ijl/D}|l\rangle\nonumber\\
&\equiv&|u_j\rangle,
\label{eq:uniform}
\end{eqnarray}
which is a uniform symmetric state. This map has been proposed by one of us and co-workers in Ref.~\cite{Prosser16} where it was shown that the states $\{|\beta_j(\xi)\rangle\}_{j=0}^{D-1}$ are more distinguishable (or more \emph{separated} from a geometrical point of view) than those in the set $\{|\alpha_j\rangle\}_{j=0}^{D-1}$ for any $\xi>0$, namely, $|\langle\beta_j(\xi)|\beta_k(\xi)\rangle|\leq|\langle\alpha_j|\alpha_k\rangle|$ for all $j\neq k$. Also, the distinguishability increases with $\xi$ and reach its maximum for $\xi=1$ where the states become uniform. They will be orthogonal  if the $|\alpha_j\rangle$'s are linearly independent and maximally separated for linearly dependent $|\alpha_j\rangle$'s.

\subsection{Separation of symmetric states}
The map $|\alpha_j\rangle\rightarrow|\beta_j(\xi)\rangle$ described above is probabilistic. As demonstrated in \cite{Prosser16}, the optimal success probability, $P_{\rm s}(\xi)$, for the transformation is
\begin{equation}   \label{eq:P_suc}
P_{\rm s}(\xi)=\frac{1}{(1-\xi)+\xi/(Da_{\rm min}^2)},
\end{equation}
$\forall\, j=0,\ldots,D-1$, where $a_{\rm min}=\min\{a_l\}_{l=0}^{D-1}$ is the smallest coefficient of the entangled state (\ref{eq:psi00}) and, consequently, of the symmetric state (\ref{eq:sym}). The physical implementation of this transformation is as follows: First, one applies a unitary coupling, $\hat{\mathcal{U}}_{1\rm a}(\xi)$, between the original system (1 in our protocol) and a two-dimensional ancillary system (\emph{ancilla}). Assuming their initial state to be $|\alpha_j\rangle_1|0\rangle_{\rm a}$, it will evolve as
\begin{eqnarray}
\hat{\mathcal{U}}_{1\rm a}(\xi)|\alpha_j\rangle_1|0\rangle_{\rm a}&=&\hat{A}_{\rm s}(\xi)|\alpha_j\rangle_1|0\rangle_{\rm a}+\hat{A}_{\rm f}(\xi)|\alpha_j\rangle_1|1\rangle_{\rm a}\nonumber\\
&=&\sqrt{P_{\rm s}(\xi)}|\beta_j(\xi)\rangle_1|0\rangle_{\rm a}+\sqrt{1-P_{\rm s}(\xi)}|\chi_j\rangle_1|1\rangle_{\rm a},
\label{eq:unit}
\end{eqnarray}
where $\{|0\rangle_{\rm a},|1\rangle_{\rm a}\}$ is the computational basis in the ancilla space, $\hat{A}_{\rm s}(\xi)$ and $\hat{A}_{\rm f}(\xi)$ are the Kraus operators acting on $\mathcal{H}_1$ (sub)space and responsible for the success and the failure in the mapping, respectively. They are given by \cite{Prosser16} 
\begin{equation}
\hat{A}_{\rm s}(\xi)=\sum_{n=0}^{D-1}\sqrt{\frac{1-\xi+\xi/Da_n^2}{1-\xi+\xi/Da_{\rm min}^2}}|n\rangle\langle n|,
\end{equation}
\begin{equation}   \label{eq:A_f}
\hat{A}_{\rm f}(\xi)=\sum_{n=0}^{D-1}\sqrt{\frac{\xi}{D}\frac{1/a_{\rm min}^2-1/a_n^2}{1-\xi+\xi/Da_{\rm min}^2}}|n\rangle\langle n|,
\end{equation}
and satisfy $\hat{A}_{\rm s}^\dag(\xi)\hat{A}_{\rm s}(\xi)+\hat{A}_{\rm f}^\dag(\xi)\hat{A}_{\rm f}(\xi)=\hat{I}$, where $\hat{I}$ is the identity operator in this space. Finally, $|\chi_j\rangle$ is the state which $|\alpha_j\rangle$ is transformed in case of a failed mapping and is given by
\begin{equation}   \label{eq:chi_st}
|\chi_j\rangle=\sum_{l=0}^{D-1}\sqrt{\frac{a_l^2-a^2_{\rm min}}{1-Da^2_{\rm min}}}e^{2\pi ijl/D}|l\rangle.
\end{equation}
These states are also symmetric but less distinguishable than the $|\alpha_j\rangle$'s, i.e., $|\langle\chi_j|\chi_k\rangle|\geq|\langle\alpha_j|\alpha_k\rangle|$ for all $j\neq k$. This is because they are restricted to a subspace of the $D$-dimensional space where the $|\alpha_j\rangle$'s live, as at least one of the coefficients in (\ref{eq:chi_st}) vanishes. Note that there is no dependence of the failure states $|\chi_j\rangle$ with the distinguishability parameter $\xi$.

\subsection{Bob's decoding}
\label{sec:sep_dec}

Let us assume that Bob applies the map $|\alpha_j\rangle\rightarrow|\beta_j(\xi)\rangle$ on system 1 to assist his decoding process of Alice's message. After the unitary coupling (\ref{eq:unit}), he performs a projective measurement on the ancilla. A projection onto $|0\rangle_{\rm a}$ ($|1\rangle_{\rm a}$), with probability $P_{\rm s}(\xi)$ [$1-P_{\rm s}(\xi)$], maps $|\alpha_j\rangle$ into $|\beta_j(\xi)\rangle$ ($|\chi_j\rangle$) and the process is successful (unsuccessful). In this section we consider the following scenario: If Bob's mapping succeeds, he accomplishes the decoding with a final measurement on system 1; otherwise, he does nothing in it, retrieving no information of its state. This is equivalent to the approach presented in Refs.~\cite{Hao00,Pati05}. In the next section we also show an improvement on this.

\begin{figure}[tbp]
\centerline{\includegraphics[width=.75\textwidth]{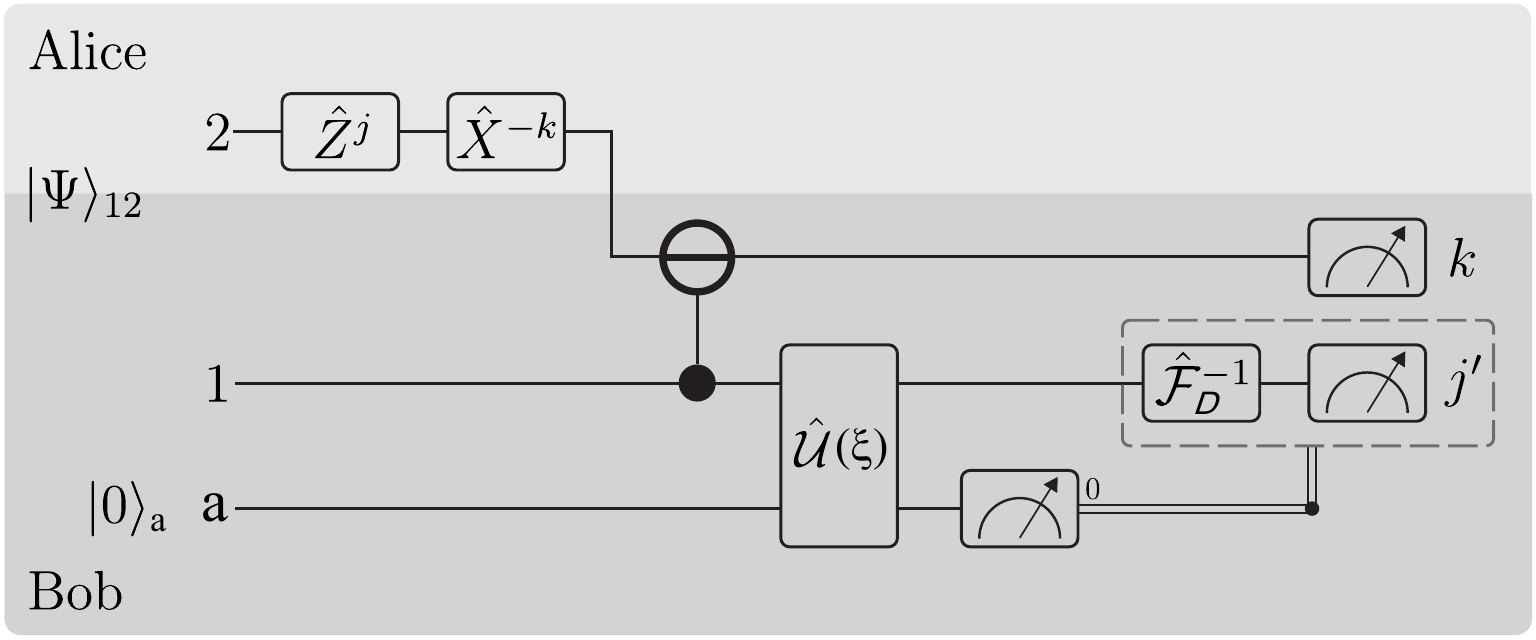}}
\caption{\label{fig:circ_sep} Probabilistic dense coding with the decoding assisted by quantum state-separation. $\hat{\mathcal{U}}_{1\rm a}(\xi)$ is given by Eq.~(\ref{eq:unit}). This circuit corresponds to a successful event (i.e., ancilla projected onto $|0\rangle$) in the transformation $|\alpha_j\rangle\rightarrow|\beta_j(\xi)\rangle$.}
\end{figure}

After a successful event, Bob applies the ME measurement given by Eq.~(\ref{eq:Pi_ME}) in order to determine, in an optimal way, the state $|\beta_j(\xi)\rangle$ and thus infer the $j$ component of Alice's message $(j,k)$. The dense coding protocol with this whole decoding procedure is illustrated in the circuit of Fig.~\ref{fig:circ_sep}. From Eqs.~(\ref{eq:Pi_ME}) and (\ref{eq:beta_st}), the probability of obtaining an outcome $\omega_l$ if the state of system 1 is $|\beta_j(\xi)\rangle$ will be
\begin{eqnarray}   
p_{jl}(\xi)&=&|\langle \mu_l|\beta_j(\xi)\rangle|^2 \nonumber\\
&=&\frac{1}{D}\sum_{m,n=0}^{D-1}b_m(\xi)b_n(\xi)\exp[2\pi i(m-n)(j-l)/D],
\label{eq:Pjl_SEP}
\end{eqnarray}
where $b_n(\xi)$ is given by (\ref{eq:b_xi}). Using Eqs.~(\ref{eq:IM_general})  and (\ref{eq:Entrop_cond}) and taking into account both successful and failed events in Bob's mapping, the mutual information between him and Alice, as a function of the distinguishability parameter $\xi$, will be
\begin{equation}   \label{eq:IM_Sep}
[I(M;R)(\xi)]^{\rm Sep} = P_{\rm s}(\xi)\underbrace{\left[\log_2 D + \frac{1}{D} \sum_{j,l=0}^{D-1} p_{jl}(\xi) \log_2  p_{jl}(\xi)\right]+\log_2 d_2} _{\displaystyle[I(M;R)(\xi)]^{\rm Sep}_{\rm suc}}.
\end{equation}
From this expression, we can obtain the boundary cases: If no transformation $|\alpha_j\rangle\rightarrow|\beta_j(\xi)\rangle$ is performed, i.e., if $\xi=0$, then $P_{\rm s}(\xi)=1$ and $[I(M;R)(0)]^{\rm Sep}=[I(M;R)]^{\rm ME}$ [see Eq.~(\ref{eq:IM_ME})], that corresponds to the results of Refs.~\cite{Barenco95,Hausladen96}. If $\xi=1$, Bob performs the maximum separation $|\alpha_j\rangle\rightarrow|u_j\rangle$ [see Eq.~(\ref{eq:uniform})] and $P_{\rm s}(\xi)=Da^2_{\rm min}$, so that $[I(M;R)(1)]^{\rm Sep}$ recovers the results of Refs.~\cite{Hao00,Pati05}. Finally, if there is no entanglement, then $P_{\rm s}(\xi)=0$ and $[I(M;R)(\xi)]^{\rm Sep}=\log_2 d_2$, as expected.

The selected term in Eq.~(\ref{eq:IM_Sep}), $[I(M;R)(\xi)]^{\rm Sep}_{\rm suc}$, corresponds to the mutual information between Alice and Bob considering only successful events in his mapping. The descending ordering of the quantities shown in (\ref{eq:IM_ME}) and (\ref{eq:IM_Sep}) reads as
\begin{equation}   \label{eq:IM_order}
[I(M;R)(\xi)]^{\rm Sep}_{\rm suc}\geq[I(M;R)]^{\rm ME}\geq[I(M;R)(\xi)]^{\rm Sep},
\end{equation}
where the equality holds only for $P_{\rm s}(\xi)=1$ or $P_{\rm s}(\xi)=0$.
The first inequality can be easily perceived if one recalls that a successful mapping transforms the input states $|\alpha_j\rangle$ into the more distinguishable ones $|\beta_j(\xi)\rangle$, thus reducing the conditional entropy (\ref{eq:Entrop_cond}) and, consequently, increasing the mutual information [see Eq.~(\ref{eq:IM_general})]. The second inequality is a consequence that the ME measurement provides, on average, a larger probability of correctly identifying the states $|\alpha_j\rangle$ than any other probabilistic strategy \cite{Jimenez11}, as the one described in this section. Therefore, it produces a larger mutual information between Alice and Bob. Despite that, the probabilistic strategies allow for Bob to identify---with a certain success probability---Alice's non-PD messages more confidently than would be possible by applying the ME strategy.

\begin{figure}[tbp]
\centerline{\includegraphics[width=1\textwidth]{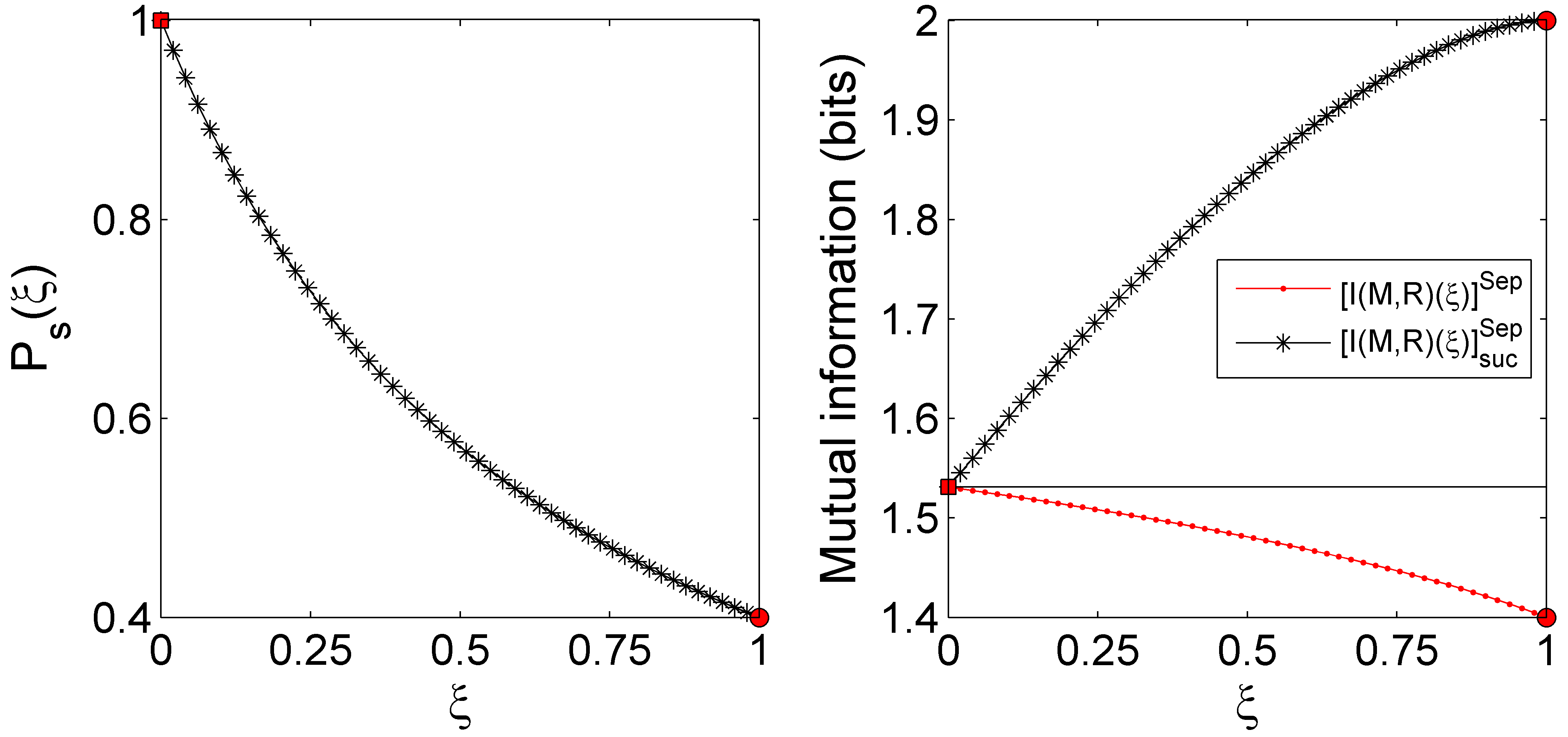}}
\caption{\label{fig:grafs_sep} Dense coding assisted by state separation using the entangled state (\ref{eq:Psi_ex1}) as resource. Success probability (\textit{left panel}) and mutual information (\textit{right panel}) as functions of the distinguishability parameter $\xi$. The \textit{squares} (\textit{circles}) plotted for $\xi=0$ ($\xi=1$) correspond to the results of Refs.~\cite{Barenco95,Hausladen96} (Refs.~\cite{Hao00,Pati05}). The \textit{solid line} in the \textit{right panel} corresponds to $[I(M;R)]^{\rm ME}$, given by Eq.~(\ref{eq:IM_ME}).}
\end{figure}

The dense coding assisted by quantum-state separation introduced here enables Bob to trade-off between the level of confidence he wishes to identify Alice's non-PD messages and the optimal success probability, $P_{\rm s}(\xi)$ [Eq.~(\ref{eq:P_suc})], for doing so. This is done by fixing the distinguishability parameter $\xi$, which sets the operations given by Eqs.~(\ref{eq:unit})--(\ref{eq:A_f}) he must implement for achieving his goal. Let us illustrate the protocol with a simple example: Assume that Alice and Bob share a two-qubit non-maximally entangled state given by 
\begin{equation}   \label{eq:Psi_ex1}
|\Psi\rangle_{12}=\sqrt{0.2}|0\rangle_1|0\rangle_2+\sqrt{0.8}|1\rangle_1|1\rangle_2.
\end{equation}
Using Eqs.~(\ref{eq:b_xi}), (\ref{eq:P_suc}), (\ref{eq:Pjl_SEP}), and (\ref{eq:IM_Sep}), we plot in Fig.~\ref{fig:grafs_sep} the success probability and the mutual information as functions of $\xi$. The solid line in the right panel corresponds to the mutual information $[I(M;R)]^{\rm ME}$, given by Eq.~(\ref{eq:IM_ME}), reached when Bob applies the ME measurement. The whole discussion above can be appreciated from these graphs. In particular, the right panel demonstrates the relationship between the mutual informations established in Eq.~(\ref{eq:IM_order}). Also, the squares plotted for $\xi=0$ and the circles for $\xi=1$ correspond to the results of the dense coding protocols proposed in Refs.~\cite{Barenco95,Hausladen96} and \cite{Hao00,Pati05}, respectively.

\section{Multistage decoding}
\label{sec:multi-stage}

The second class of probabilistic dense coding we propose here is based on the following: After a failed attempt in the map $|\alpha_j\rangle\rightarrow|\beta_j(\xi)\rangle$ described in the previous section, the state of system 1, $|\alpha_j\rangle$, is transformed into $|\chi_j\rangle$ [see Eq.~(\ref{eq:chi_st}) and subsequent discussion]. These failure states are restricted to a $D'$-dimensional subspace of $\mathcal{H}_1$, where $D'=D-\mu$, and $\mu$ denotes the multiplicity of the smallest Schmidt coefficient, namely $a_{\rm min}$. The case $\mu=D$ is not of interest because the entangled state is uniform and the protocol works perfectly. For $\mu=D-1$, the failure states are identical (up to a global phase) for all $j=0,\ldots,D-1$ and hence, retrieve no information about the $|\alpha_j\rangle$'s. On the other hand, for $1\leq\mu\leq D-2$ we have $2\leq D'\leq D-1$, and the failure states, although less distinguishable than the $|\alpha_j\rangle$'s, will still carry some information about them. Therefore, instead of simply doing nothing in case of failure in the mapping (as considered in Sect.~\ref{sec:sep_dec} and also in Ref.~\cite{Pati05}), Bob may try to identify the state $|\chi_j\rangle$ (and hence $|\alpha_j\rangle$) to infer $j$ in Alice's message $(j,k)$. He can apply any strategy that discriminates, optimally, among the equally likely symmetric states in the set $\{|\chi_j\rangle\}_{j=0}^{D-1}$, having in mind that these states are now linearly dependent. Depending on the strategy he adopts and other features of the entangled state (e.g., $D$ and the multiplicity of the other Schmidt coefficients), Bob can iterate this process in as many stages as allowed by the quantum channel. This is the essence of the multistage decoding which, as we will see, increases the mutual information between Alice and Bob. Note that this scheme applies \emph{only} when none of the systems are qubits and the condition $D\geq 3$ holds.

\subsection{Decoding with maximum-confidence quantum measurements}

\begin{figure}[tbp]
\centerline{\includegraphics[width=.8\textwidth]{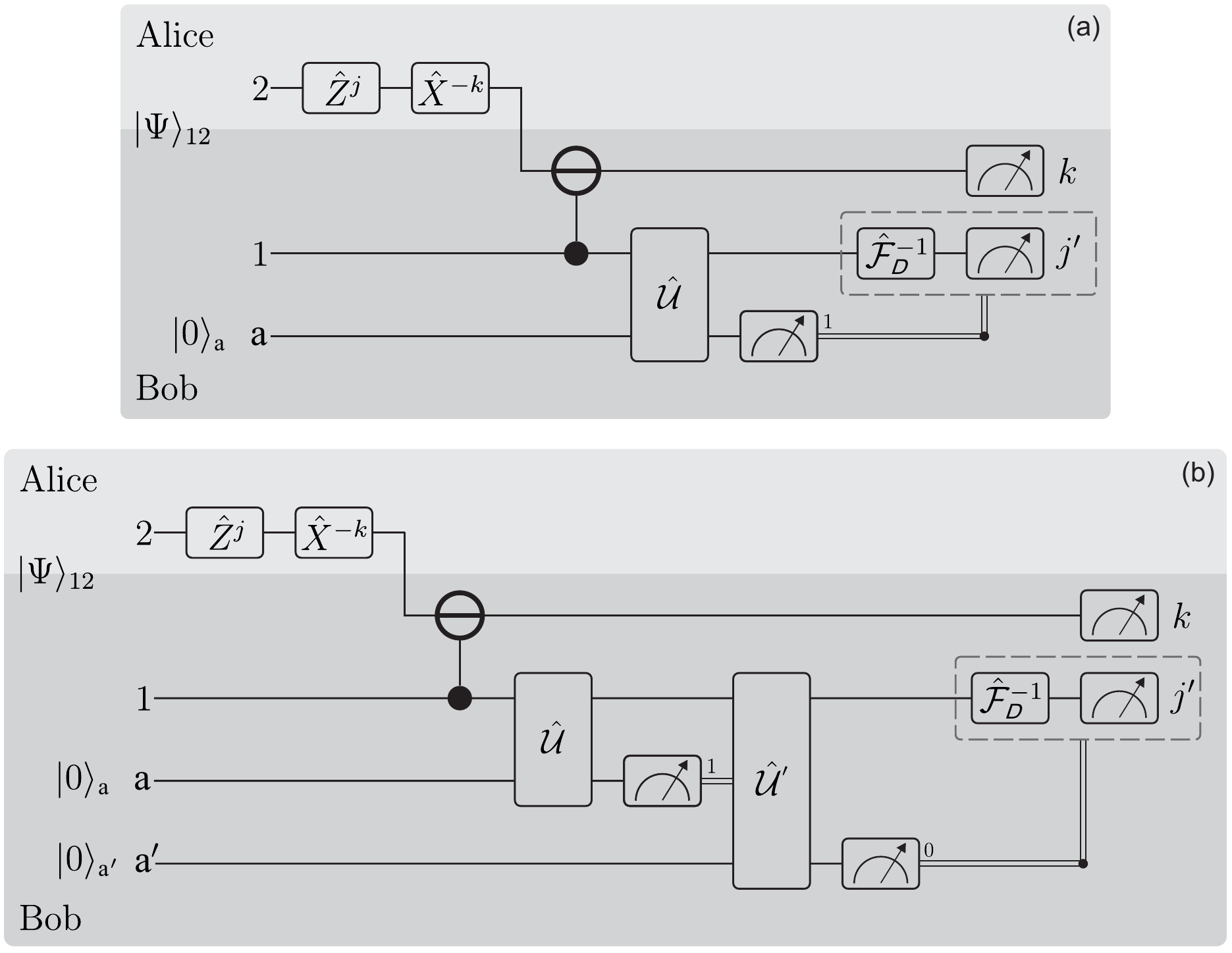}}
\caption{\label{fig:circ_seq} Probabilistic dense coding with the circuits showing a two-stage decoding after a failure in the first attempt of the MC measurement (ancilla projected onto $|1\rangle$). $\hat{\mathcal{U}}$ is given by (\ref{eq:uniform}) with $\xi=1$. \textbf{a} Second stage performed with the ME measurement. \textbf{b} Second stage performed with the MC measurement again (the circuit shows a successful attempt).}
\end{figure}

We will describe the multistage decoding considering that at any stage Bob implements quantumstate separation, he sets the distinguishability parameter at its maximum,\footnote{This is not a requirement for the process but will be adopted here in order to establish a comparison with previous results in the literature and also to make the whole discussion clearer.} i.e., $\xi=1$ (thus, from now on we omit $\xi$ from the corresponding expressions). For equally likely symmetric states, this map followed by a ME measurement (\ref{eq:Pi_ME}) on the successfully transformed states discriminate among them with the maximum confidence \cite{Jimenez11}. In the present context, the linearly independent states $|\alpha_j\rangle$ (\ref{eq:sym}) are transformed in the uniform states $|u_j\rangle$ (\ref{eq:uniform}) which will be orthogonal. Accordingly, they will be discriminated unambiguously (which means confidence 1) with the optimal success probability~\cite{Chefles98}
\begin{equation}    \label{eq:P_UD}
P_{\rm s,\mathnormal{1}}=Da_{\rm min}^2,
\end{equation}
calculated from Eq.~(\ref{eq:P_suc}), where the subindex $\mathnormal{1}$ denotes the first stage in the decoding process. This is the probability that Bob will identify perfectly the non-PD messages from Alice. Using Eqs.~(\ref{eq:IM_general}) and (\ref{eq:Entrop_cond}),  the mutual information between them will be 
\begin{eqnarray}
[I(M;R)]^{\rm MC,\mathcal{S}} & = & P_{\rm s,\mathnormal{1}}[I(M;R)]^{\rm MC}_{\rm suc,\mathnormal{1}}+(1-P_{\rm s,\mathnormal{1}})
[I(M;R)]_{\rm fail,\mathnormal{1}}^{\mathcal{S}} \nonumber\\
&=&Da_{\rm min}^2\log_2d_2D +(1-Da_{\rm min}^2)[I(M;R)]_{\rm fail,\mathnormal{1}}^{\mathcal{S}},
\label{eq:IM_seq}
\end{eqnarray}
where $[I(M;R)]_{\rm fail,\mathnormal{1}}^{\mathcal{S}}$ is the mutual information when Bob applies a discrimination strategy $\mathcal{S}$ to identify the state $|\chi_j\rangle$ [Eq.~(\ref{eq:chi_st})] resulting from a failure in the mapping $|\alpha_j\rangle\rightarrow|u_j\rangle$. For instance, if he applies the ME measurement (\ref{eq:Pi_ME}) one obtains
\begin{equation}   \label{eq:IM_ME_fail}
[I(M;R)]_{\rm fail,\mathnormal{1}}^{\rm ME}=\log_2d_2D+\frac{1}{D} \sum_{j,l=0}^{D-1} |\langle \mu_l|\chi_j\rangle|^2 \log_2  |\langle \mu_l|\chi_j\rangle|^2 .
\end{equation}
This case is illustrated in the quantum circuit of Fig.~\ref{fig:circ_seq}a. On the other hand, by applying the MC measurement and a given strategy $\mathcal{S}$ in case of failure, this mutual information will be
\begin{eqnarray}
[I(M;R)]_{\rm fail,\mathnormal{1}}^{\rm MC,\mathcal{S}} & = & P_{\rm s,\mathnormal{2}}
[I(M;R)]_{\rm suc,\mathnormal{2}}^{\rm MC}  +(1-P_{\rm s,\mathnormal{2}})[I(M;R)]_{\rm fail,\mathnormal{2}}^{\mathcal{S}} ,
\label{eq:IM_seq2}
\end{eqnarray}
where
\begin{equation}   \label{eq:P_suc_2}
P_{\rm s,\mathnormal{2}}=(D-\mu)\min_{a_l\neq a_{\rm min}}\left\{\frac{a_l^2-a^2_{\rm min}}{1-Da^2_{\rm min}}\right\}_{l=0}^{D-1},
\end{equation}
which is the product between the dimension of the subspace spanned by $\{|\chi_j\rangle\}$ and the square of the smallest nonvanishing coefficient of these states \cite{Jimenez11}. Additionally,
\begin{equation}   \label{eq:IM_suc_2}
[I(M;R)]_{\rm suc,\mathnormal{2}}^{\rm MC}=\log_2d_2D+\frac{1}{D} \sum_{j,l=0}^{D-1} |\langle \mu_l|u'_j\rangle|^2 \log_2  |\langle \mu_l|u'_j\rangle|^2 ,
\end{equation}
where $\{|u'_j\rangle\}_{j=0}^{D-1}$ is a set of uniform symmetric states in a $(D-\mu)$-dimensional subspace of $\mathcal{H}_1$, resulting from a successful mapping $|\chi_j\rangle\rightarrow|u'_j\rangle$ in the second stage of the decoding process. They are maximally distinguishable but not orthogonal since the failure states $|\chi_j\rangle$ are linearly dependent. Thus, $|\langle \mu_l|u'_j\rangle|^2 \neq\delta_{jl}$ and $[I(M;R)]_{\rm suc,\mathnormal{2}}^{\rm MC}<[I(M;R)]_{\rm suc,\mathnormal{1}}^{\rm MC}$. Figure~\ref{fig:circ_seq}b shows the quantum circuit with a well succeeded MC measurement at the second stage of the decoding process.

The same reasoning for $[I(M;R)]_{\rm fail,\mathnormal{1}}^{\mathcal{S}}$ in (\ref{eq:IM_seq}) can be applied to $[I(M;R)]_{\rm fail,\mathnormal{2}}^{\mathcal{S}}$ in (\ref{eq:IM_seq2}) as long as the failure states in the second stage still carry information about the $|\alpha_j\rangle$'s. This process may then be iterated under this latter condition. Bob's discrimination strategy on a given stage will depend on what is advantageous for him to infer Alice's message. For instance, if his concern is to infer this message more confidently, it might be possible (as we will see in the example below) that a successful MC measurement on that stage still provides higher confidence than a ME measurement at the first stage. In this scenario, the success probability for his task will increase by employing the multistage decoding.

Equation~(\ref{eq:IM_seq}) tells us that whenever
\begin{equation}   \label{eq:condition}
[I(M;R)]_{\rm fail,\mathnormal{1}}^{\mathcal{S}}>\log_2d_2,
\end{equation}
the mutual information $[I(M;R)]^{\rm MC,\mathcal{S}}$ will always be larger than the one achieved by discarding the failure outcome in the decoding, as can be seen by comparing it with Eq.~(\ref{eq:IM_Sep}) for $\xi=1$. The condition (\ref{eq:condition}) is fulfilled in both cases discussed above, either by applying the ME measurement [see Eq.~(\ref{eq:IM_ME_fail})] or another round of the MC measurement [see Eqs.~(\ref{eq:IM_seq2})--(\ref{eq:IM_suc_2})]. Therefore, the multistage optimal discrimination strategies enable to increase the transmission rate of classical information in the probabilistic dense coding protocols as compared with previous schemes \cite{Wu06,Pati05} which did not explored this unique feature of qudit states.

\subsection{Example}
\begin{figure}[tbp]
\centerline{\includegraphics[width=1\textwidth]{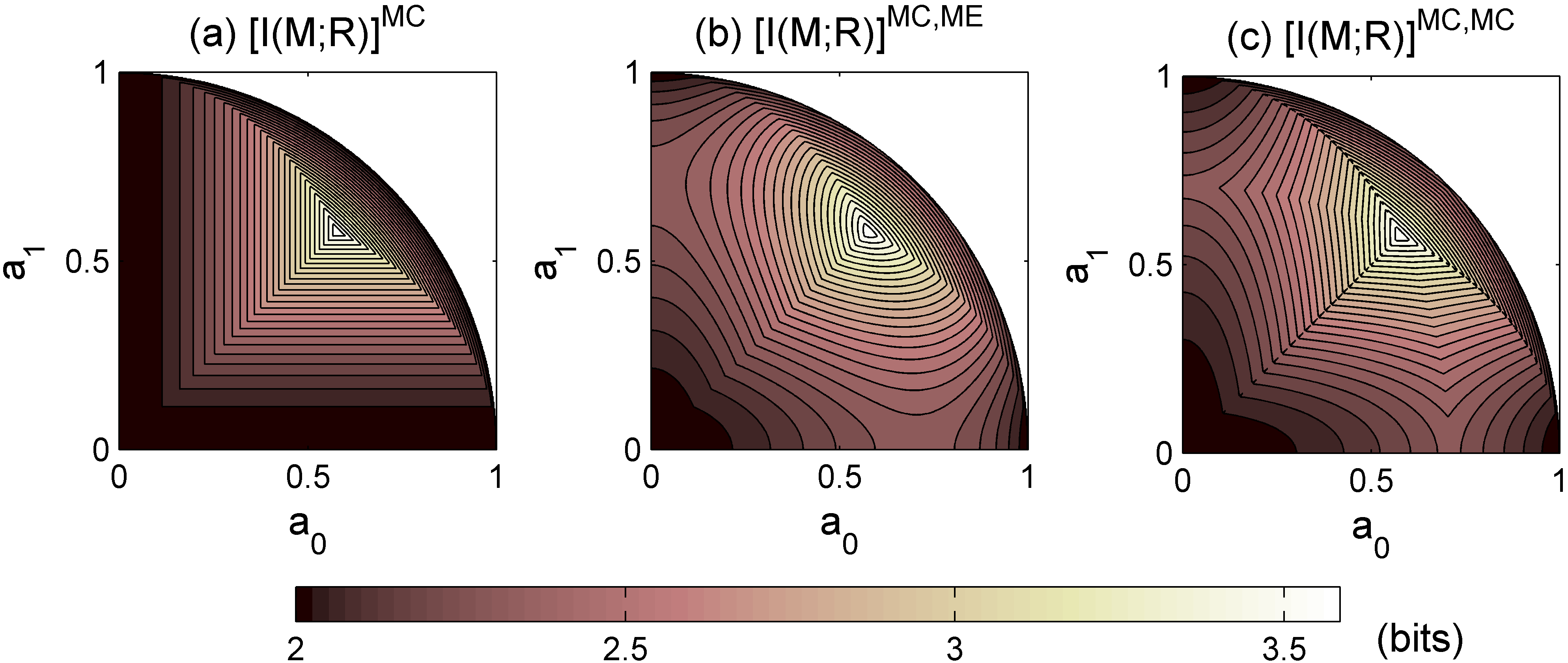}}
\caption{\label{fig:IM_seq}  Mutual information in the dense coding as  function of the Schmidt coefficients $a_0$ and $a_1$, for an entangled state with $d_1=3$ and $d_2=4$.  The decoding is performed with \textbf{a} single-stage MC measurement, \textbf{b} MC measurement followed by a ME measurement in case of failure in the first stage, and \textbf{c} two-stage MC measurement.}
\end{figure}

We return to the example of Sect.~\ref{sec:ME} to illustrate the main results presented in this section. For an entangled state with $d_1=3$ and $d_2=4$, Bob can perform a two-stage decoding. Figure~\ref{fig:IM_seq} shows the specified mutual informations as functions of the Schmidt coefficients $a_0$ and $a_1$ ($a_2=\sqrt{1-a_0^2-a_1^2}$). In Fig.~\ref{fig:IM_seq}a we plot $[I(M;R)]^{\rm MC}$ which is obtained from Eq.~(\ref{eq:IM_Sep}) with $\xi=1$. It corresponds to a single-stage decoding: Bob performs a MC measurement where nothing else is done in case of failure. The results of two-stage decodings are shown in Figs.~\ref{fig:IM_seq}b, c. The plot of Fig.~\ref{fig:IM_seq}b corresponds to the mutual information $[I(M;R)]^{\rm MC,ME}$ where a ME measurement is performed at the second stage. It is obtained from Eqs.~(\ref{eq:IM_seq}) and (\ref{eq:IM_ME_fail}). Finally, in Fig.~\ref{fig:IM_seq}c we plot the mutual information $[I(M;R)]^{\rm MC,MC}$ for a two-stage MC measurement, which is obtained from Eqs.~(\ref{eq:IM_seq}) and (\ref{eq:IM_seq2})--(\ref{eq:IM_suc_2}). Clearly, both two-stage decodings increase the  mutual information between Alice and Bob.

\begin{figure}[tbp]
\centerline{\includegraphics[width=.7\textwidth]{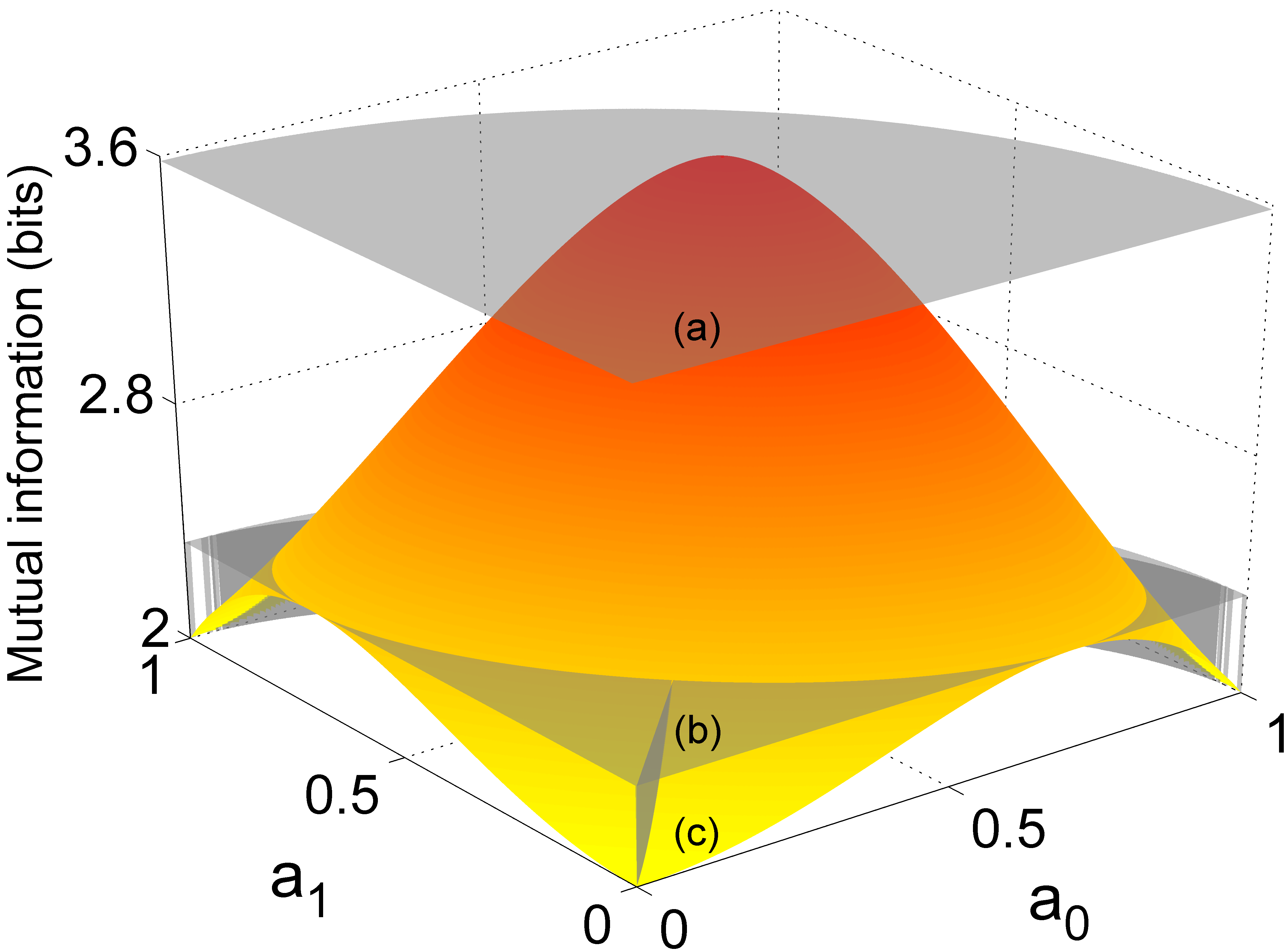}}
\caption{\label{fig:IM_SMC} Mutual information in the dense coding as  function of the Schmidt coefficients $a_0$ and $a_1$, for an entangled state with $d_1=3$ and $d_2=4$.  The surfaces \textit{a} and \textit{b} correspond to the mutual information after successful events in the first and second stages, respectively, of a two-stage MC measurement; \textit{a} is obtained from the underbraced part of Eq.~(\ref{eq:IM_Sep}) with $\xi=1$ and \textit{b} from Eq.~(\ref{eq:IM_suc_2}). The surface \textit{c} is reproduced from Fig.~\ref{fig:IM_ME} and corresponds to the mutual information of a single-stage decoding via ME measurement.}
\end{figure}

\begin{figure}[tbp]
\centerline{\includegraphics[width=1\textwidth]{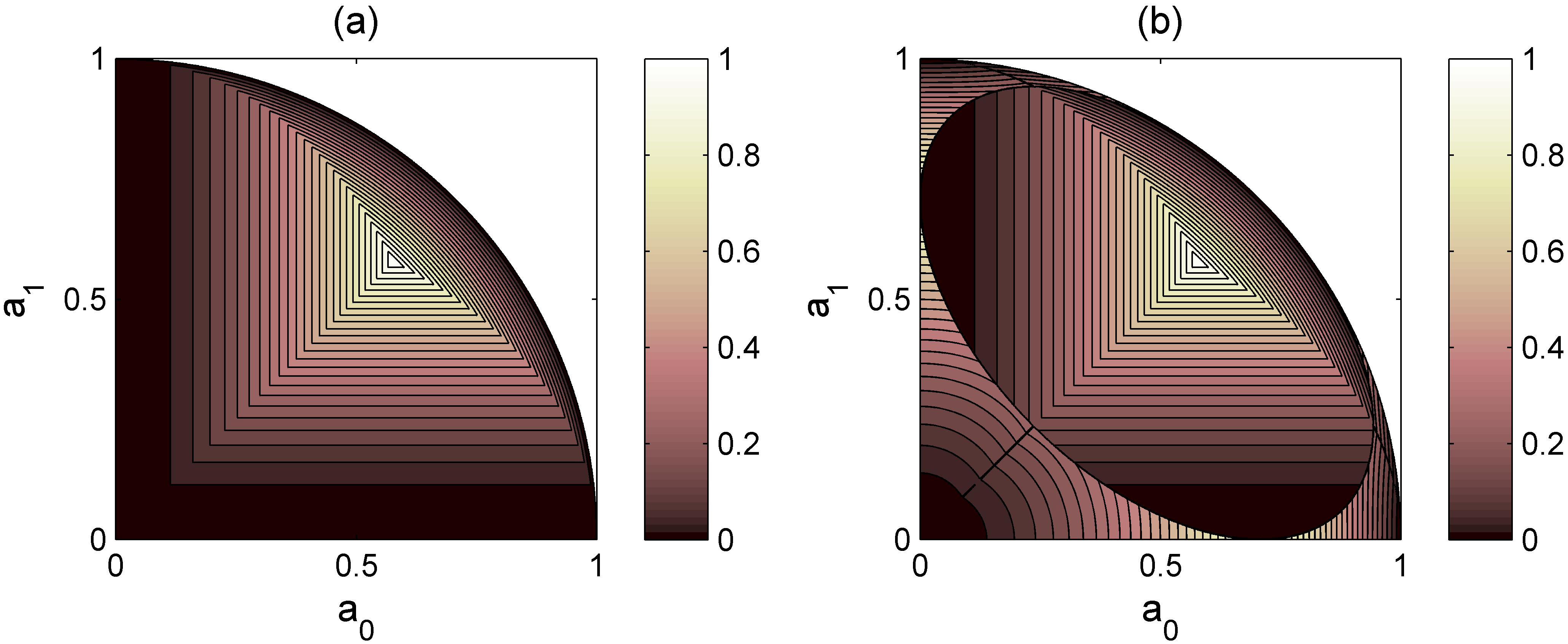}}
\caption{\label{fig:prob_seq} Success probabilities which Bob will decode Alice's messages with a larger confidence than the one achieved with the ME measurement. They are shown as function of the Schmidt coefficients $a_0$ and $a_1$, for an entangled state with $d_1=3$ and $d_2=4$. \textbf{a} Single-stage and \textbf{b} two-stage decoding with MC measurements.}
\end{figure}

Another interesting feature of this process can be appreciated in the graphs of Fig.~\ref{fig:IM_SMC}. Considering the decoding with a two-stage MC measurement, we plot the mutual informations considering only successful events in the first [surface (a)] and second [surface (b)] stages, obtained from the underbraced part of Eq.~(\ref{eq:IM_Sep}) (with $\xi=1$) and Eq.~(\ref{eq:IM_suc_2}), respectively. For the sake of comparison, we also reproduce, in surface (c), the mutual information $[I(M;R)]^{\rm ME}$ of Fig.~\ref{fig:IM_ME}, obtained from a single-stage decoding performed with a ME measurement. Surface (a) shows that the first inequality in Eq.~(\ref{eq:IM_order}) is fulfilled. In this case, the mutual information reaches the maximum value provided by a maximally entangled channel, since the non-PD messages are distinguished with full confidence, although only probabilistically. This is independent of the amount of entanglement shared between Alice and Bob. In case of failure in the first stage of the MC measurement and a well succeeded event in the second one, the corresponding mutual information is greatly reduced as seen from surface (b). However, aside few exceptions,\footnote{In surface (b), the curves that reach the lower bound of 2~bits correspond to entangled states whose minimum Schmidt coefficient has multiplicity two, so that the second stage of MC measurement would be useless.} it is larger than the lower bound of 2~bits that would be achieved if Bob had simply given up of inferring this part of Alice's message after a failure in the first stage. What is more interesting to observe is that the regions of surface (c) below of surface (b) indicate the range of entangled states for which a success in the second stage of the MC measurement leads Bob to infer Alice's message still more confidently than would be with a single-stage decoding via ME measurement. Therefore, for such entangled states, Bob increases the success probability of identifying the received non-PD messages with more confidence. This is shown in the graphs of Fig.~\ref{fig:prob_seq}. Figure~\ref{fig:prob_seq}a shows the success probability in the first stage of the MC measurement, which is the probability of reaching the mutual information shown in the surface (a) of Fig.~\ref{fig:IM_SMC}. Figure~\ref{fig:prob_seq}b shows the overall success probability of performing the decoding with a larger confidence than with ME measurement. The improvement in the region below surface (b) of Fig.~\ref{fig:IM_SMC} is clear.

\section{Decoding schemes applied to quantum key distribution}
\label{sec:QKD}

The decoding schemes presented in Sects.~\ref{sec:ME}, \ref{sec:separation}, and \ref{sec:multi-stage} may also have impact in high-dimensional quantum key distribution (QKD) protocols. Here, we will address, qualitatively, which would be such possible impacts---a thorough and quantitative analysis is beyond the scope of the present paper. Hopefully, this will open the doors for further investigations on the subject. 

Let us consider a generalization to high dimensions of the two-state QKD protocol developed by Bennett \cite{Bennett92-qkd} (the B92 protocol). In order to generate a secure shared key with Bob, Alice sends him a quantum system prepared with equal probability in one of the $D$-dimensional symmetric states $\{|\alpha_j\rangle\}_{j=0}^{D-1}$, given by Eq.~(\ref{eq:sym}), each one corresponding to the logical ``dit'' $j=0,\ldots,D-1$. Bob discriminates unambiguously (confidence 1) among the possible states with the optimal success probability given by Eq.~(\ref{eq:P_UD}). Then, over a public classical channel, he tells Alice whether the discrimination succeeded or not. If it has succeeded, they keep the dit; if it has failed, they discard it. After many repetitions of this procedure, Alice and Bob will share a sequence of $j$'s which will be used as a key. 

Now, assume that an eavesdropper (Eve) intercepts the quantum system before it arrives to Bob. As the states are not orthogonal, she cannot determine perfectly which one Alice has prepared. What she can do is to apply a given discrimination strategy to infer the state in the best possible way. Then, she keeps the inferred dit, prepares a new state identical to the inferred one, and send it to Bob. Obviously, this procedure will introduce errors in the protocol since Eve will, sometimes, send the wrong dit to Bob. To detect these errors, Alice and Bob compare some of their dits in a public channel. If there are no errors, there is no eavesdropper and the remaining dits will be kept to form the key. Otherwise, if there are errors within a given range, the presence of the eavesdropper will be very likely and, in this case, they will discard all the dits and repeat the process.

The error rate introduced by Eve in this procedure as well as how much she learns about the key will depend on the discrimination strategy she adopts. In order to remain undetected (or, at least, hard to be detected), she must introduce as few errors as possible. In this scenario, let us analyze, qualitatively, the advantages and shortcomings of the strategies studied here. First, we consider the two extreme cases where Eve applies either ME or MC strategy:
\begin{itemize}
\item[(i)] If Eve chooses to apply the ME strategy (Sect.~\ref{sec:ME}), she will not get a perfect key. However, she introduce the least amount of errors. If these errors are below the unavoidable errors from other sources (e.g., imperfect communication channel and detection), she will remain undetected. 
\item[(ii)] If Eve chooses to apply MC strategy (Sect.~\ref{sec:separation} for $\xi=1$), she will learn the key with much more confidence. However, since the overall probability of identifying the states incorrectly is always larger than in the ME strategy, she will introduce a higher error rate. Thus, she could be detected more easily.  
\end{itemize}
How the discrimination strategies studied in Sects.~\ref{sec:separation} and \ref{sec:multi-stage} would fit in this problem? First of all, since both are probabilistic, the errors in identifying the states are always larger than in the deterministic ME strategy \cite{Jimenez11}. Accordingly, the error rate introduced by Eve would always be higher than in (i) should she applies any of those strategies. However, there may still be advantages in using them:
\begin{itemize}
\item[(iii)] If Eve chooses to apply the state separation  (Sect.~\ref{sec:separation} for $0<\xi<1$) followed by a ME measurement, she can trade-off (by setting the distinguishability parameter $\xi$) between the level of confidence in learning the key and the error rate introduced in the protocol. The former would be larger than in (i) and the latter, lower than in (ii). 
\item[(iv)] If Eve chooses to apply the multistage decoding (Sect.~\ref{sec:multi-stage}), she will have the chance of learning the key with as much confidence as in (ii) and, at the same time, to introduce a lower error rate. This is because the overall error probability in identifying the states is reduced in comparison with a single-stage MC measurement. 

\end{itemize}

Therefore, for high-dimensional QKD---considering the B92 protocol using symmetric states---Alice and Bob must be aware of these many discrimination strategies which Eve can implement to get the key. With this knowledge, they will be able to establish the range of errors introduced by a possible eavesdropper adopting any of these strategies and hence, to detect his/her presence. 

\section{Conclusion}
\label{sec:conc}
In summary, we revisited the problem of dense coding with non-maximally entangled states for protocols adopting the standard encoding method. Using the recent advances in the field of quantum-state discrimination, we proposed optimal probabilistic decoding schemes which generalized and improved over previous approaches. Firstly, we demonstrated that by applying quantum-state separation onto the information carriers, one can controllably increase the distinguishability of the messages, with an optimal success probability. This scheme was shown to include those of Refs.~\cite{Barenco95,Hausladen96} and \cite{Hao00,Pati05} as special cases and continuously interpolate between them, enabling the decoder to trade-off between the level of confidence desired to identify the received messages and the success probability for doing so. After that we introduced the concept of multistage decoding, based on successive  attempts of state discrimination when there is a failure in the current attempt. We showed that this scheme applies only for dense coding with qudits and is advantageous over previous reports \cite{Pati05,Wu06} as it increases the mutual information between the sender and receiver. Finally, we discussed, qualitatively, the application of these decoding schemes in a high-dimensional QKD protocol. Further, quantitative studies on this subject may bring advances in this field. \\

\noindent \textbf{Acknowledgements} This work was supported by the Brazilian agencies CNPq through Grant No.~485401/2013-4 and  FAPEMIG through Grant No.~APQ-00240-15.

\section*{Appendix}
In this appendix we show that the mutual information between Alice and Bob in the dense coding protocol described in Sect.~\ref{sec:QDC} is given by Eqs.~(\ref{eq:IM_general}) and (\ref{eq:Entrop_cond}). For a pair of random variables $M$ and $R$, the mutual information is defined as \cite{CoverBook}
\begin{equation}   \label{eq:IM_app}
I(M;R)=H(M)-H(M|R),
\end{equation}
where $H(M)$ and $H(M|R)$ are the Shannon and conditional entropies, respectively. In the present context, the random variable $M$ is associated with the set of possible messages $\{(j,k)|j=0,\ldots,D-1;k=0,\ldots,d_2-1\}$ encoded by Alice in the quantum state $\hat{\rho}_{jk}=|\alpha_j\rangle\langle\alpha_j|\otimes|k\rangle\langle k|$ [see Eq.~(\ref{eq:Message_state})] with a probability $p(\hat{\rho}_{jk})=1/(d_2D)$. Thus, the Shannon entropy will be given by 
\begin{eqnarray}
H(M)&=&-\sum_{j=0}^{D-1}\sum_{k=0}^{d_2-1}p(\hat{\rho}_{jk})\log_2p(\hat{\rho}_{jk}) \nonumber\\
&=&\sum_{j=0}^{D-1}\sum_{k=0}^{d_2-1}\frac{1}{d_2D}\log_2\frac{1}{d_2D}\nonumber\\
&=&\log_2d_2D.
\label{eq:Ent_Shannon_app}
\end{eqnarray}
On the other hand, the random variable $R$ is associated with the set of Bob's measurement results $\{\omega_{lm}|l=0,\ldots,N-1;m=0,\ldots,d_2-1\}$ in his decoding process. The index $l$ ($m$) is connected with the result of a measurement on system 1 (2), and the number of values it assumes, $N$ ($d_2$), will be explained later. With these definitions, the conditional entropy will be written as
\begin{equation}    \label{eq:Ent_cond_app}
H(M|R)=-\sum_{j=0}^{D-1}\sum_{l=0}^{N-1}\sum_{k,m=0}^{d_2-1}p(\omega_{lm})
p(\hat{\rho}_{jk}|\omega_{lm})\log_2p(\hat{\rho}_{jk}|\omega_{lm}).
\end{equation}
Using Bayes' rule
\begin{equation}
p(\hat{\rho}_{jk}|\omega_{lm})=p(\omega_{lm}|\hat{\rho}_{jk})\frac{p(\hat{\rho}_{jk})}{p(\omega_{lm})},
\end{equation}
we can rewrite (\ref{eq:Ent_cond_app}) as
\begin{equation}    \label{eq:Ent_cond_app2}
H(M|R)=-\sum_{j=0}^{D-1}\sum_{l=0}^{N-1}\sum_{k,m=0}^{d_2-1}p(\hat{\rho}_{jk})
p(\omega_{lm}|\hat{\rho}_{jk})\log_2\left[p(\omega_{lm}|\hat{\rho}_{jk})
\frac{p(\hat{\rho}_{jk})}{p(\omega_{lm})}\right],
\end{equation}
where $p(\omega_{lm})$ is the overall probability of obtaining the outcome $\omega_{lm}$, which, from the symmetry of the problem \cite{Jimenez11}, is given by
\begin{equation}
p(\omega_{lm})=\frac{1}{d_2D};
\end{equation}
$p(\omega_{lm}|\hat{\rho}_{jk})$ is the conditional probability of obtaining the outcome $\omega_{lm}$ given that the prepared state was $\hat{\rho}_{jk}$. As we showed in Sect.~\ref{sec:QDC}, the state of system 2 is perfectly discriminated by a projective measurement onto the computational basis $\{|m\rangle\langle m|\}_{m=0}^{d_2-1}$. However, for discriminating the states of system 1, we need, in general, to implement an $N$-outcome generalized measurement $\{\hat{\Pi}_l^{\mathcal{S}}\}_{l=0}^{N-1}$, with $N\geq D$, where ${\mathcal{S}}$ stands for the strategy to be adopted. Therefore, we obtain 
\begin{eqnarray}
p(\omega_{lm}|\hat{\rho}_{jk})&=&{\rm Tr}(\hat{\rho}_{jk}\hat{\Pi}_{lm})\nonumber\\
&=&{\rm Tr}\left[\left(|\alpha_j\rangle\langle\alpha_j|\otimes|k\rangle\langle k|\right)\left(\hat{\Pi}_l^{\mathcal{S}}\otimes|m\rangle\langle m|\right)\right]
\nonumber\\
&=&{\rm Tr}(\hat{\rho}_{j}\hat{\Pi}_l^{\mathcal{S}})\delta_{km}.
\end{eqnarray}
Replacing these results in Eq.~(\ref{eq:Ent_cond_app2}), the conditional entropy will be given by
\begin{eqnarray}
[H(M|R)]^{\mathcal{S}}&=&-\frac{1}{d_2D}\sum_{j=0}^{D-1}\sum_{l=0}^{N-1}\sum_{k,m=0}^{d_2-1}
{\rm Tr}(\hat{\rho}_{j}\hat{\Pi}_l^{\mathcal{S}})\delta_{km}\log_2{\rm Tr}(\hat{\rho}_{j}\hat{\Pi}_l^{\mathcal{S}}) \nonumber\\
&=&-\frac{1}{D}\sum_{j=0}^{D-1}\sum_{l=0}^{N-1}
{\rm Tr}(\hat{\rho}_{j}\hat{\Pi}_l^{\mathcal{S}})\log_2{\rm Tr}(\hat{\rho}_{j}\hat{\Pi}_l^{\mathcal{S}}) \nonumber\\
&=&[H(M_1|R_1)]^{\mathcal{S}},
\label{eq:Ent_cond_app_3}
\end{eqnarray}
where $[H(M_1|R_1)]^{\mathcal{S}}$ is the conditional entropy associated with the encoding/decoding process for the part of the message encoded in system 1. Using the results of Eqs.~(\ref{eq:Ent_Shannon_app}) and (\ref{eq:Ent_cond_app_3}) in Eq.~(\ref{eq:IM_app}), we obtain the mutual information of Eq.~(\ref{eq:IM_general}), which ends the proof.

\end{document}